**Should univariate Cox regression be used for feature selection with respect to time-to-event outcomes?**

Rong Lu, PhD*

*The Quantitative Sciences Unit, Division of Biomedical Informatics Research, Department of Medicine, Stanford University, Stanford, California

**Corresponding Author:** Rong Lu, PhD (https://orcid.org/0000-0003-4321-9144); ronglu@stanford.edu; 3180 Porter Drive, Palo Alto, CA 94304.

**Conflicts of Interest:** None

**Funding Sources:** This work is partially supported by the Biostatistics Shared Resource (BSR) of the NIH-funded Stanford Cancer Institute: P30CA124435. As part of the Stanford Center for Clinical and Translational Research and Education, Rong Lu also received support from UL1-TR003142.

**Role of the Funder/Sponsor:** The funders had no role in the design and conduct of the study; collection, management, analysis, and interpretation of the data; preparation, review, or approval of the manuscript; and decision to submit the manuscript for publication.

**KEY POINTS**

**Question:** Should univariate Cox regression be used for feature selection with respect to time-to-event outcomes?

**Findings:** When features are independent and the true models are multivariate Cox proportional hazards models, Gaussian regression of log-transformed survival time (response variable) with only two covariates outperformed both the univariate Cox proportional hazards model and logistic regression in feature selection, in terms of not only higher sensitivity, comparable specificity, but also higher accuracy of effect size ranking, regardless of the sample size and censoring rate values.

**Meaning:** This study demonstrates the importance of including Gaussian regression of log-transformed survival time in feature selection practice for time-to-event outcomes.

**ABSTRACT**

**IMPORTANCE:** Time-to-event outcomes are commonly used in clinical trials and biomarker discovery studies and have been primarily analyzed using Cox proportional hazards models. But it's unclear which statistical models should be recommended for feature selection tasks when time-to-event outcomes are of the primary interest.

**OBJECTIVE:** To explore if Gaussian regression of log-transformed survival time could outperform Cox proportional hazards models in feature selection.

**DESIGN:** In this simulation study, the true models are multivariate Cox proportional hazards models with 10 covariates. For all feature selection comparisons, it's assumed that only 5 out the 10 true features are observed/measured for all model fitting, along with 5 random noise features. Each sample size and censoring rate scenario is explored using 10,000 simulation datasets. Different statistical models are applied to the same dataset to estimate feature effects. Model performance is compared using sensitivity, specificity, and accuracy of effect size ranking.

**RESULTS:** When features are independent and the true models are multivariate Cox proportional hazards models, Gaussian regression of log-transformed survival time (response variable) with only two covariates outperformed both the univariate Cox proportional hazards model and logistic regression in feature selection, in terms of not only higher sensitivity, comparable specificity, but also higher accuracy of effect size ranking, regardless of the sample size and censoring rate values.

**CONCLUSIONS AND RELEVANCE:** This study demonstrates the importance of including Gaussian regression of log-transformed survival time in feature selection practice for time-to-event outcomes.

**INTRODUCTION**

Time-to-event outcomes are commonly used in clinical trials and biomarker discovery studies and have been primarily analyzed using Cox proportional hazards models [1-4]. But it's still unclear which statistical models should be recommended for feature selection tasks when time-to-event outcomes are of the primary interest.

The main goal of feature selection analysis is to identify features/covariates that can affect the outcome of interest and estimate/rank their effect sizes. One of the major challenges in feature selection tasks is that there is no way to guarantee that all true features are measured and available for analysis, especially for biomarker discovery studies [2-5]. This challenge directly leads to the difficulty of designing simulation studies for comparing statistical models' performance in feature selection tasks. This is because all simulation studies need to specify a true model to generate simulation datasets, but it's unclear how to choose true models to produce findings with good generalizability in practice settings. As a result, I believe that simulation studies for comparing feature selection methods should aim for helping test/correct current believe of best practice, instead of aiming for increasing the generalizability of the findings. That is, I believe that the choice of the true model only needs to reflect the assumptions made for methods that are currently believed to be the best in the field.

Based on biomarker studies published in recent years on high-impact journals, it seems that the multivariate Cox proportional hazards model is still generally believed to be the best model for estimating biomarkers' effects on time-to-event outcomes, regardless of the censoring rate [1-4]. Even when other method, such as univariate Cox regression or Log-rank test or LASSO-regularized Cox regression, is used for feature selection many researchers choose to report the effect sizes estimated from multivariate Cox model as the main finding [6-8]. Therefore, I believe it's important to compare model performance for feature selection in simulation studies where multivariate Cox proportional hazards model are used as the true models to generate survival time.

In this simulation study, four models' performance in feature selection tasks will be explored under the assumptions that: 1) the time-to-event outcome values are generated from multivariate Cox proportional hazards models with different correlation levels between features, different sample sizes, and different censoring rates, and 2) not all true features are measured.

**METHODS**

In this simulation study, the true models are always multivariate Cox proportional hazards models with 10 true features/covariates for all simulation scenarios, but with different choices of parameter values for baseline cumulative risk function, feature correlation levels, sample sizes, and censoring rates.

Parameters for generating survival time:

Each simulation dataset is created using different parameter values for baseline cumulative risk function. These parameters are randomly drawn using R code *sample(c(10:200)/5,1),* which was designed to reduce the probability of generating survival/follow-up time values that are too concentrated towards zero and have extremely long right tails at the same time. The 10 true features are generated using multivariate Gaussian distribution with mean=0 and variance=1 for all features. The covariance values of multivariate Gaussian will differ between simulation scenarios, to create either independent true features or highly correlated true features. The effect sizes of the 10 true features are always 1, 2, 3, …, 10.  Each sample size (n = 500, 1000, 2000, or 5000) and censoring rate scenario (10% or 50%) is explored using 10,000 simulation datasets.

Design to reflect the assumption that not all true features are measured:

In all simulation scenarios, it's assumed that only 5 out of the 10 true features are observed/measured, along with 5 noise features that are generated independently from the true features. The 5 observed true features are randomly picked with equal probability. Those noise features are generated using the same multivariate Gaussian that is used to generate true features.

Models of interest:

The following regression models are applied to every simulation dataset to screen the 10 available features (5 true features + 5 noise features) and rank their effect sizes:

1. Multivariate Cox proportional hazards model: to explore the up bound of performance statistics, I assumed that, in ideal situation, the field experts already identified the 5 true features, and the research goal was to test/confirm their hypothesis. Therefore, only the 5 true features are included in fitting one multivariate Cox models for each simulation dataset.
2. Univariate Cox proportional hazards model: 10 univariate Cox models are fitted for each simulation dataset, one for each available feature.
3. Logistic regression: 10 Logistic regression models are fitted for each simulation dataset, one for each available feature. The response variable is the event indicator. The 2 covariates are the feature under screening and the survival/follow-up time.
4. Gaussian regression: 10 Gaussian regression models are fitted for each simulation dataset, one for each available feature. The response variable is the log-transformed survival time. The 2 covariates are the feature under screening and the event indicator.

For all models, a feature is selected if its coefficient p-value <0.05 and the effect size rankings are based on coefficient p-values as well (smaller p indicates bigger effect).

Assessment of Model Performance:

Model performance is assessed using sensitivity, specificity, and accuracy of effect size ranking. For the calculation of all performance statistics, each simulation dataset is considered as one analysis unit. That is, the sample size for performance assessment is always 10,000 for each simulation scenario. For the calculation of sensitivity and specificity, each analysis unit is considered positive-by-test only if all true features are correctly identified and is considered negative-by-test only if all noise features are correctly identified. Similarly, for the calculation of accuracy of effect size ranking, each analysis unit is considered having accurate ranking only if all true features are correctly ranked.

To help clarify all methodology details and ensure reproducibility of this work, all scripts and scripts' outputs are provided as attachments and are documented in Supplementary Table 1. These scripts can also be modified to facilitate simulation studies of other parameter/model choices, which might be needed to better reflect the survival time distribution, feature correlation level, sample size, and censoring rate of observational datasets that are collected in practice.

**RESULTS**

The main results of this simulation study are summarized in Table 1.

When features are independent

Gaussian regression (of log-transformed survival time with two covariates) outperformed both the univariate Cox proportional hazards model and the logistic regression in feature selection tasks, in terms of not only higher sensitivity, comparable specificity, but also higher accuracy of effect size ranking, regardless of the sample size and censoring rate choices.

Based on this observation, I believe that both the univariate Cox proportional hazards model and the logistic regression are not preferred for screening features or estimating effect sizes when the true model is believed to be a multivariate Cox regression. Instead, the following analysis steps are recommended if censoring rate $\geq 50\%$:

- Step 1: fit the Gaussian regression model to test each feature one at a time, using the log-transformed survival/follow-up time as the response variable and using the feature and the event indicator as the only two covariates.
- Step 2: select the feature if the p-value of feature coefficient reported from Gaussian model is less than 0.05
- Step 3: rank all selected features using the p-values of feature coefficients reported by the Gaussian model. This ranking can be used as the main result of effect size ranking.

- Step 4: (Sensitivity analysis) fit a multivariate Cox model using all features selected by Gaussian regression, then rank all features using feature coefficients' p-values reported by this multivariate Cox model

When features are highly correlated

Gaussian regression (of log-transformed survival time with two covariates) outperformed all other 3 models in terms of perfect sensitivity and comparable specificity. But its effect size ranking accuracy is far from acceptable, especially when the sample size is large. The univariate Cox model also seems to have very low accuracy of effect size ranking when the sample size is large. The logistic regression has the worst sensitivity and very low accuracy of effect size ranking as well.

Based on this observation, I believe that both the univariate Cox proportional hazards model and the logistic regression are still not the best choices for screening features or estimating effect sizes when the true model is believed to be a multivariate Cox proportional hazards model. And the Gaussian model can no longer be trusted for effect size ranking. The following 3 analysis steps are recommended when features are highly correlated:

- Steps 1 & 2 are the same as the recommendations for independent feature selection.
- Step 3: fit a multivariate Cox model using all features selected by Gaussian regression, then rank all features using feature coefficients' p-values reported by this multivariate Cox model.

**Table 1:** Summary of performance statistics (n=10,000)

| Sample Size | Censoring Rate | True Feature Correlation | Model | Feature Selection Sensitivity | Feature Selection Specificity | Accuracy of Effect Size Ranking |
|---|---|---|---|---|---|---|
| 500 | 10% | 0 | Multivariate Cox | 56.91% | 76.66% | 62.24% |
| | | | Univariate Cox | 37.65% | 76.92% | 51.42% |
| | | | Logistic Regression | 0% | 77.00% | 0.74% |
| | | | Gaussian Regression | 45.89% | 77.48% | 60.43% |
| 1000 | 10% | 0 | Multivariate Cox | 71.80% | 77.31% | 78.73% |
| | | | Univariate Cox | 55.59% | 77.33% | 70.41% |
| | | | Logistic Regression | 0% | 77.53% | 0.82% |
| | | | Gaussian Regression | 64.72% | 77.09% | 78.60% |
| 2000 | 10% | 0 | Multivariate Cox | 87.64% | 77.65% | 91.22% |
| | | | Univariate Cox | 73.71% | 77.73% | 85.30% |
| | | | Logistic Regression | 0% | 77.43% | 0.92% |
| | | | Gaussian Regression | 81.25% | 77.13% | 91.77% |
| 5000 | 10% | 0 | Multivariate Cox | 98.18% | 76.90% | 82.89% |
| | | | Univariate Cox | 92.64% | 77.11% | 97.91% |
| | | | Logistic Regression | 0% | 77.99% | 1.06% |
| | | | Gaussian Regression | 97.45% | 76.45% | 99.45% |
| 500 | 50% | 0 | Multivariate Cox | 41.67% | 75.78% | 46.82% |
| | | | Univariate Cox | 21.69% | 76.13% | 35.53% |
| | | | Logistic Regression | 0% | 77.41% | 0.80% |
| | | | Gaussian Regression | 45.31% | 76.17% | 60.19% |
| 1000 | 50% | 0 | Multivariate Cox | 58.23% | 77.14% | 64.99% |
| | | | Univariate Cox | 39.33% | 77.40% | 54.16% |
| | | | Logistic Regression | 0% | 77.63% | 0.91% |
| | | | Gaussian Regression | 64.77% | 77.07% | 78.60% |
| 2000 | 50% | 0 | Multivariate Cox | 74.87% | 77.50% | 80.81% |
| | | | Univariate Cox | 58.20% | 77.70% | 71.15% |
| | | | Logistic Regression | 0% | 77.67% | 0.76% |
| | | | Gaussian Regression | 81.16% | 77.27% | 91.77% |
| 5000 | 50% | 0 | Multivariate Cox | 92.56% | 77.74% | 93.60% |
| | | | Univariate Cox | 80.26% | 77.78% | 90.63% |
| | | | Logistic Regression | 0% | 77.60% | 0.86% |
| | | | Gaussian Regression | 97.47% | 76.47% | 99.46% |
| 500 | 10% | 0.8 | Multivariate Cox | 84.27% | 79.25% | 34.74% |
| | | | Univariate Cox | 98.88% | 87.38% | 10.72% |
| | | | Logistic Regression | 0.56% | 87.38% | 0.76% |
| | | | Gaussian Regression | 100% | 86.42% | 27.05% |
| 1000 | 10% | 0.8 | Multivariate Cox | 86.10% | 79.65% | 49.37% |
| | | | Univariate Cox | 99.09% | 87.64% | 17.22% |
| | | | Logistic Regression | 0.50% | 87.61% | 0.94% |
| | | | Gaussian Regression | 100% | 87.20% | 1.58% |
| 2000 | 10% | 0.8 | Multivariate Cox | 87.85% | 80.17% | 61.23% |
| | | | Univariate Cox | 99.10% | 88.19% | 1.95% |
| | | | Logistic Regression | 0.76% | 87.29% | 0.69% |
| | | | Gaussian Regression | 100% | 87.52% | 0.80% |
| 5000 | 10% | 0.8 | Multivariate Cox | 89.15% | 80.45% | 31.92% |
| | | | Univariate Cox | 98.92% | 88.24% | 1.50% |
| | | | Logistic Regression | 0.66% | 87.21% | 0.70% |
| | | | Gaussian Regression | 100% | 86.91% | 0.90% |
| 500 | 50% | 0.8 | Multivariate Cox | 81.53% | 78.02% | 24.13% |
| | | | Univariate Cox | 96.87% | 87.02% | 5.62% |
| | | | Logistic Regression | 0.51% | 87.17% | 0.79% |
| | | | Gaussian Regression | 100% | 86.35% | 26.77% |
| 1000 | 50% | 0.8 | Multivariate Cox | 83.47% | 78.34% | 37.10% |
| | | | Univariate Cox | 96.98% | 87.48% | 10.14% |
| | | | Logistic Regression | 0.48% | 87.21% | 0.77% |
| | | | Gaussian Regression | 100% | 87.26% | 1.57% |
| 2000 | 50% | 0.8 | Multivariate Cox | 84.66% | 79.18% | 50.47% |
| | | | Univariate Cox | 97.23% | 87.25% | 16.89% |
| | | | Logistic Regression | 0.82% | 87.51% | 0.86% |
| | | | Gaussian Regression | 100% | 87.46% | 0.80% |
| 5000 | 50% | 0.8 | Multivariate Cox | 85.88% | 79.30% | 60.02% |
| | | | Univariate Cox | 97.03% | 87.56% | 1.47% |
| | | | Logistic Regression | 0.62% | 87.20% | 0.86% |
| | | | Gaussian Regression | 100% | 86.93% | 0.90% |

**DISCUSSION**

This study demonstrates the importance of including Gaussian regression of log-transformed survival time in feature selection practice for time-to-event outcomes. But it's still unclear whether the recommendations made in this work could be further improved by incorporating more assumption-checking steps or using other types of models [9-10].

All scripts shared in this work also include sections to explore performance of R function cox.zph(), for checking the proportional hazards (PH) assumption. When features are independent, it's observed that cox.zph() tends to report more severe PH assumption violation for true features that actually have bigger effect sizes, in both the univariate and multivariate Cox regression analysis, which might be a little supersizing and counter intuitive (Sup Fig 1). It's also observed that cox.zph() often fails to identify PH assumption violation for noise features. The corresponding performance statistics are provided in attached scripts' outputs. When features are highly correlated, cox.zph() can fail in execution. The corresponding error messages are provided in scripts as comments. Well-designed simulation studies of cox.zph() might be needed before using it to guide model choice between regular Cox regression and its variants with time-varying covariates or time-varying coefficients [11].

# SUPPLEMENTARY TABLES AND FIGURES

**Supplementary Table 1:** simulation_scripts.xlsx

**Supplementary Figure 1:** Under the assumptions made in this simulation study, R function cox.zph() tends to report more severe proportional hazards assumption violation for true features that have bigger effect sizes.

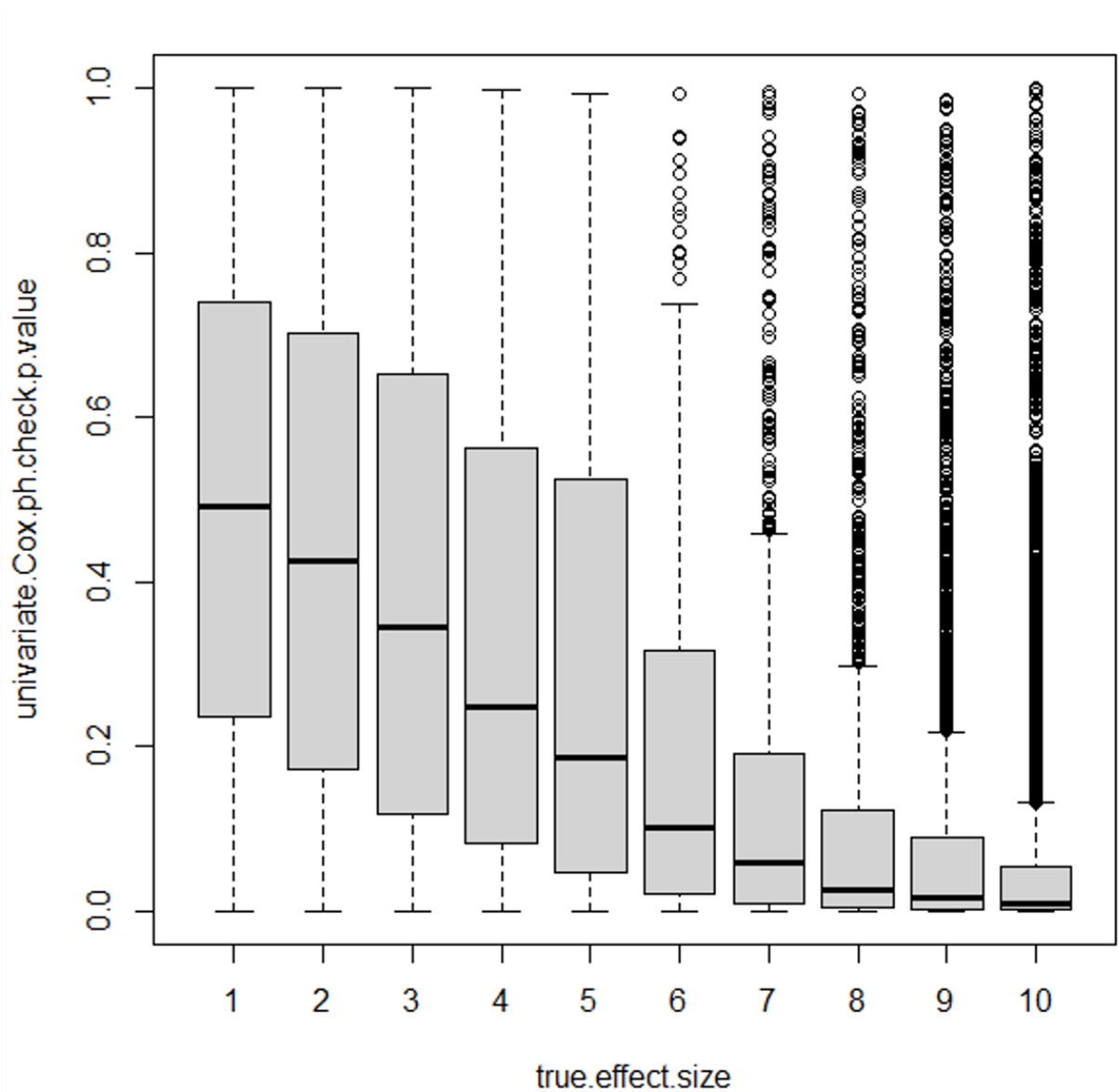